\documentclass[prb,twocolumn,aps,superscriptaddress,showpacs,floatfix]{revtex4}
\usepackage{amsmath}
\usepackage{amssymb}
\usepackage{graphicx}

\begin{document}

\title{
Current-Voltage Characteristics of Weyl Semimetal Semiconducting Devices, Veselago Lenses and Hyperbolic Dirac Phase
}

\author{R. D. Y. Hills}
\affiliation{Department of Physics, Loughborough University, Leicestershire, LE11 3TU, United Kingdom}

\author{A. Kusmartseva}
\affiliation{Department of Physics, Loughborough University, Leicestershire, LE11 3TU, United Kingdom}

\author{F. V. Kusmartsev}
\affiliation{Department of Physics, Loughborough University, Leicestershire, LE11 3TU, United Kingdom}

\date{\today}

\begin{abstract}
The current-voltage characteristics of a new range of devices built around Weyl semimetals has been predicted using the Landauer formalism. The potential step and barrier have been re-considered for a three-dimensional Weyl semimetals, with analogies to the two-dimensional material graphene and to optics.  With the use of our results we also show how a Veselago lens can be made from Weyl semimetals, e.g. from NbAs and NbP. Such a lens may have many practical applications and can be used as a probing tip in a scanning tunneling microscope (STM). The ballistic character of Weyl fermion transport inside the semimetal tip, combined with the ideal focusing of the Weyl fermions (by Veselago lens) on the surface of the tip may create a very narrow electron beam from the tip to the surface of the studied material. With a Weyl semimetal probing tip the resolution of the present STMs can be improved significantly, and one may image not only individual atoms but also individual electron orbitals or chemical bonding and therewith to resolve the long-term issue of chemical and hydrogen bond formation.

We show that applying a pressure to the Weyl semimental, having no centre of spacial inversion one may model matter at extreme conditions such as those arising in the vicinity of a black hole. As the materials Cd$_{3}$As$_{2}$ and Na$_{3}$Bi show an asymmetry in their Dirac cones, a scaling factor was used to model this asymmetry. The scaling factor created additional regions of no propagation and condensed the appearance of resonances. We argue that under an external pressure there may arise a topological phase transition in Weyl semimetals, where the electron transport changes character and becomes anisotropic. There a hyperbolic Dirac phases occurs where there is a strong light absorption and photo-current generation.
\end{abstract}

\maketitle

\pagenumbering{arabic}
\pagestyle{plain}


\section{Introduction}
	The recent influx of study into the two-dimensional material graphene has highlighted the field of gapless semiconductors. In graphene the conduction and valence bands meet at a point known as a Dirac point \cite{b1} and mass-less charge carriers follow a linear energy-momentum relation \cite{b11}. Due to the linear dispersion relation, high Fermi-velocity ($v_{f}\sim c/300$) and massless charge carriers, graphene quasiparticles can be modelled by the relativistic Dirac equation \cite{b12}.

	The results obtained in graphene are equally applicable to a broad class of materials generally named topological insulators \cite{b2,b3}.  In Ref.\cite{b2} it was shown that on the interface between the two insulating semiconductors CdTe and HgTe(Se) an inverted band structure may arise. The inverted band structure creates a metallic conducting layer associated with the Dirac gapless spectrum. The single Dirac point is protected by a time reversal symmetry and the conductivity in the Dirac point at zero temperature tends to infinity.

	Three-dimensional materials such as Ag$_{2}$Se and Ag$_{2}$Te have been shown to act as small gap semiconductors \cite{b23,b24} and Ag$_{2}$Te can experience a phase transition from narrow-gap semiconductor to gapless semiconductor with a linear spectrum \cite{b25}. Other materials such as grey tin \cite{b26} and mercury telluride \cite{b27} have been shown to possess zero-gap properties with a parabolic dispersion relation. In the case of mercury telluride the size of the energy gap can be adjusted by replacing atoms of mercury with the lighter element cadmium. With a specific concentration of cadmium the dispersion relation becomes gap-less and linear \cite{b28}.
		
	Recently, Na$_{3}$Bi has been shown to act with a three-dimensional linear dispersion relation \cite{b29}. The crystal structure of Na$_{3}$Bi forms a hexagonal Brillouin zone in the $k_{x}-k_{y}$ plane similar to the two-dimensional material graphene. This forms three-dimensional Dirac cones close to the center of the Brillouin zone.
		
	The three-dimensional Dirac cones have also been shown in the material Cd$_{3}$As$_{2}$ \cite{b34,p2,b35,b36}. This material possessed a non-symmetrical Dirac cone \cite{b34} in the $k_{y}-k_{z}$ direction which is slightly shifted from the Fermi level \cite{p2} and a Fermi-velocity 1.5 that of graphene \cite{b35}. These qualities make Cd$_{3}$As$_{2}$ a good candidate for the exploration of Weyl semimetals and three-dimensional Dirac cones.

	It has been predicted that a new family of Weyl semimetals including TaAs \cite{c1,c2}, TaP, NbAs \cite{c3}, and NbP may possess as many as 12 pairs of Weyl points in the Brillouin zone \cite{p3}. Similarly to Cd$_{3}$As$_{2}$ the Weyl points appear slightly shifted from the Fermi energy; the Weyl points on the $k_{z}=0$ plane appearing at about 2 meV above, and the off-plane Weyl points appearing about 21 meV below the Fermi level.


\section{ Types of Topological Semimetals and Their Origins}
	In the last couple of years, four types of topological semimetals have been described \cite{Bernevig2016,Kane2016,Nayak2017}. The first class is the three-dimensional Dirac semimetals. In this case there is four fold degeneracy of the Dirac cones. The degeneracy is related to the time reversal symmetry or T invariance (the  two-fold Kramers degeneracy) and the valley or spatial inversion symmetry or P invariance, which provides also the additional two fold degeneracy. 

	The second class are so-called Weyl semimetals (WSM). They arise when one of these symmetries is broken and the degeneracy is lifted. Therefore, the Weyl semimetal has two-fold degenerate bands arising on two different Dirac cones, which always come in pairs. These two Dirac cones have different topological charge describing the chirality 
of the Weyl fermions. On one cone the topological charge or the chirality is equal to $c=+1$, while on the other cone $c=-1$, that is the chirality on these two Dirac cones are opposite to each other. These Weyl points act like 'magnetic' monopoles or 'hedgehog' in momentum space with a charge given by their chirality. It can be well described with the use of the concept of the Berry curvature and 'Berry flux', which is flowing from one monopole into another.  

	In the third class of topological metals the touching of the valence and conduction bands in k-space occurs along one-dimensional curves forming a Dirac valley, which may be linear or in the form of a loop. In general, the bands crossing, which results in the creation of the Dirac cones or Dirac valleys may have a more complicated form. There may arise situations in solids where linear and quadratic many-fold-band crossing is possible and is stabilised by crystallographic space group symmetries. In all cases it is important to take into account together with; time-reversal T, and spatial inversion, P symmetries as well as the spin-orbit coupling, which provides both the possibility of bands crossing and the formation of Dirac cones. 

	The origin of these Weyl or Dirac points and valleys are in spin-orbital interaction, which is mixing $s$ and $p$ orbitals of heavy elements such as Sn and Pd\cite{Rashba-1960}. The spin-orbital interaction causes a band inversion of $s$ and $p$ bands, that leads to a non-trivial band topology where the valence band
has $s$ character while the conduction band has $p$ character. This is opposite to the natural order of band filling\cite{opto9}. That mechanism of Dirac point formation has been shown at the formation of the semimetalic heterojunction\cite{b2}. It was also noticed in the band structure calculations\cite{Pulci-2013,Pulci-2013A}.

	A peculiar band inversion has also been observed in topological nodal line semimetals AX$_2$ (where A = Ca, Sr, Ba and X = Si, Ge, Sn \cite{38}), in Cu$_3$PdN\cite{39} and in a topological semimetal KNa$_2$Bi \cite{40} indicating the presence of either reflection or point group symmetries in the case of non-centrosymmetric crystallographic group.


\section{Symmetries of Weyl semimetals and Transmission Through a Potential Barrier}
	Now we would like to focus on the second class of topological semimetals, where there are two Dirac 'monopoles' in the momentum space created at the Weyl points (WP) WP1 and WP2, which have opposite topological charge. With the use of the kp-method\cite{Bir-1974} one may easily show that in the vicinity of each of these points the system may be described by two Hamiltonians;  $\hat{H}_{+}=v_{f}\hat{p}\cdot\vec{\sigma}$ for the positive chirality or the monopole and $\hat{H}_{-}=-v_{f}\hat{p}\cdot\vec{\sigma}$ for negative chirality or the anti-monopole. Here each Hamiltonian is a two by two matrix and is named as the Weyl Hamiltonian \cite{p3,b50,b51,c4} where $v_{f}$ is the Fermi velocity, $\hat{p}$ is the three-dimensional momentum operator and $\vec{\sigma}=(\sigma_x,\sigma_y,\sigma_z)$ is the vector of the Pauli spin matrices. This form of $\hat{H}_{\pm}$ is similar to the Hamiltonian which may describe two-dimensional crystals such as graphene, silicene and germanene\cite{opto9} which is also a two by two matrix. However it operates in a two-dimensional momentum space $p_x,p_y$ and $\vec{{\Sigma}}=(\sigma_x,\sigma_y)$. Due to the structure of $\hat{H}=v_{f}\hat{p}\cdot\vec{\Sigma}$, the graphene buckling or the term $\Delta\sigma_z$ added to this Hamiltonian may create a gap in the graphene spectrum\cite{opto9}. In contrast, for the Weyl Hamiltonians such an extra term will not change the three-dimensional, but graphene-like, linear dispersion relation
\begin{equation}
	E=\pm v_{f}\sqrt{k_{x}^{2}+k_{y}^{2}+k_{z}^{2}}.
\end{equation}
where we can change the $k_z$ momentum by a simple shift $k_z\rightarrow k_z+\Delta$. This indicates that the Weyl Hamiltonian has additional symmetry in a comparison with the graphene Hamiltonian\cite{opto9} and this is indeed a feature of the topological stability of the Weyl point, which is simply shifted by the value $\Delta$ along $k_z$ direction due to such a perturbation. Similarly if there is an external potential $V(x)$, such that the Weyl Hamiltonian is perturbed by the term $V I$, where $I$ is the identity matrix, the Weyl point is shifted in the energy space by $V$
\begin{equation}
	E=V\pm v_{f}{|\vec{k}|},
\end{equation}
	where the $x$ dependent potential $V(x)$ has been replaced by a constant potential $V$. Thus, by applying various perturbations to the Weyl Hamiltonian we are not able to open an energy gap, which is not the case for the Dirac Hamiltonian applied to describe the gapless spectrum of graphene where an energy gap can be introduced by many means\cite{Pototsky,Yung-2013,Liu-2015}.	
	
	First of all we may notice that under the time reversal transformation T, or the spatial inversion one Weyl point is transformed into another, i.e. WP1 $\rightarrow$ WP2. In other words the topological charge of the Weyl fermions changes to the opposite one. For the penetration of a Weyl fermion through a potential step or barrier this also means that the incident wave of the Weyl fermion associated with the WP1 under the time reversal operation  $t\rightarrow -t$, is transformed into a reflection wave of the Weyl fermion associated with the other, the second WP2. Thus, due to such symmetries the transmission of the WP1 fermion incident from the left side of the step can be connected to the transmission of the WP2 fermion, but incident from the right side of the potential step. Note that during the tunneling or transmission event the topological charge of the fermion is conserved. There may exist, in principle, intervalley scattering, that is the chirality changes, but it looks as if this is not a simple process, therefore we may only study the transmission of a Weyl fermion of one fixed chirality. This simplifies the issue of tunneling for Weyl fermions. Moreover due to the similarities with the graphene Hamiltonian it is reasonable to apply the same theoretical methodologies to Weyl fermions of fixed chirality in the hope to provide graphene like properties in a three-dimensional material.

	 With the energy eigenvalues of the Hamiltonian, suitable eigenvectors can be selected. These eigenvectors can then be used as wave functions describing charge carriers in a Weyl semimetal and take the form
\begin{equation}
	\psi=
	\left[\begin{array}{ccc}
		\psi_{1}\\
		\psi_{2}
	\end{array}\right]
	=
	\left[\begin{array}{ccc}
		e^{iqx+ik_{y}y+ik_{z}z}\\
		\alpha e^{iqx+i\theta+ik_{y}y+ik_{z}z}
	\end{array}\right]
	\label{wave-functions}
\end{equation}
with
\begin{equation}
	q=\sqrt{\frac{\left(E-V\right)^{2}}{\hbar^{2}v_{f}^{2}}-k_{y}^{2}-k_{z}^{2}},
\end{equation}
\begin{equation}
	\alpha=\frac{|E-V|\sin\left(\phi\right)}{E-V+|E-V|\cos\left(\phi\right)}.
\end{equation}
It is convenient  to parametrise the momenta $q, k_y$ and $k_z$ in spherical co-ordinates as
\begin{equation}
	q=\frac{|E-V|}{\hbar v_{f}}\sin\phi \cos\theta,
\end{equation}
\begin{equation}
	k_{y}=\frac{|E-V|}{\hbar v_{f}}\sin\phi \sin\theta,
\end{equation}
\begin{equation}
	k_{z}=\frac{|E-V|}{\hbar v_{f}}\cos\phi.
\end{equation}


\section{Scattering Properties of a Weyl Semimetal Heterostructure}
\begin{figure}
	\centerline{\includegraphics[scale=0.29]{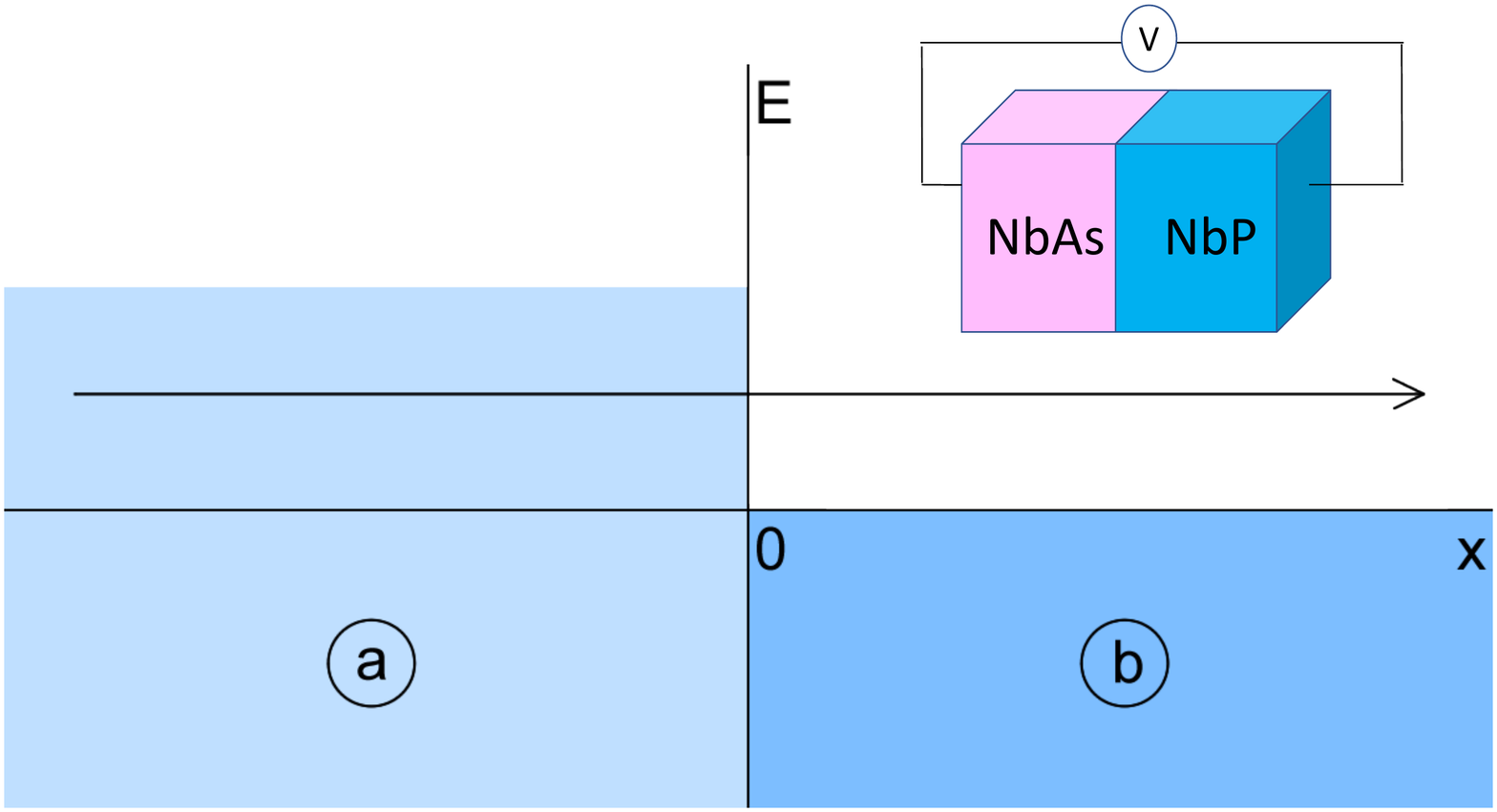}}
	\caption{Diagram of the potential step problem. The two independent regions of different Weyl semimetal materials have been labled as $a$ and $b$. This is a simple model of a heterostructure made of two Weyl semimetals, e.g. the semimetal $a$ can be NbAs while the semimetal $b$ can be NbP. An external potential $V_a$ separating them is associated with affinity or a work function of these two different materials. A potential step is placed in the $x$-direction with a heights $V_{a}\neq 0$ and $V_{b}=0$. The shaded region shows where hole transport is present.}
	\label{weyl-step}
\end{figure}
	The one-dimensional cross-section of the  potential step is presented in Fig. \ref{weyl-step}. This is a simple model of a heterostructure made of two Weyl semimetals, as an example the $a$ region semimetal could be NbAs while the $b$ region semimetal could be NbP. An external potential $V_a$ separating them is  associated with affinity or a work function of these two different materials which create the scattering system with a reflection point at $x=0$.

	In the potential step the initial and final mediums are not identical, therefore the transmission through the system cannot simply be taken to be $|t|^{2}$. Instead the expression for transmission must be obtained from the conservation of probability current \cite{b13,b40}
	\begin{equation}
		\frac{\partial}{\partial t}|\psi|^{2}+\nabla\cdot {\bf j}=0.
	\end{equation}
	As the system here is time independent only the probability current
	\begin{equation}
		{\bf j} =\psi^{*} {\bf \sigma} \psi
	\end{equation}
	needs to be considered. From the continuity equation, the probability current into the system must equal the probability current out of the system. With the wave functions in Eq.(\ref{step-psi-a}) and Eq.(\ref{step-psi-b}), this results in the following expression for transmission
\begin{equation}
	T=|t|^{2}\frac{\alpha_{b}\cos\left(\theta_{b}\right)}{\alpha_{a}\cos\left(\theta_{a}\right)}.
	\label{transmission}
\end{equation}
	
	The subscripts $a$ and $b$ correspond to the groups of constants for the corresponding region in Fig. \ref{weyl-step}, the details of how this is obtained have been included in the supplementary information\cite{Supplementary}. With the expression for the transmission probability in Eq.(\ref{transmission}) the potential step can now be solved using the wave functions in Eq.(\ref{wave-functions})
		\begin{equation}
			\psi_{a}=
			\left[\begin{array}{ccc}
				\left(e^{iq_{a}x}+re^{-iq_{a}x}\right)e^{ik_{y}y}e^{ik_{z}z}\\
				\left(\alpha_{a}e^{iq_{a}x+i\theta_{a}}-r\alpha_{a}e^{-iq_{a}x-i\theta_{a}}\right)e^{ik_{y}y}e^{ik_{z}z}
			\end{array}\right]
			\label{step-psi-a}
		\end{equation}
		where the incident and reflected components have been included. The subscript $a$ corresponds to region $a$ in Fig. \ref{weyl-step}. The wave functions on the right of the step (corresponding to region $b$ in Fig. \ref{weyl-step}) only contain a transmitted component and is therefore given as
		\begin{equation}
			\psi_{b}=
			\left[\begin{array}{ccc}
				te^{iq_{b}x}e^{ik_{y}y}e^{ik_{z}z}\\
				t\alpha_{b}e^{iq_{b}x+i\theta_{b}}e^{ik_{y}y}e^{ik_{z}z}
			\end{array}\right].
			\label{step-psi-b}
		\end{equation}

	Then at the interface located at $x=0$, continuity of the wave functions require that $\psi_{a}=\psi_{b}$. Solving these simultaneous equations, with the transmission probability in Eq.(\ref{transmission}) results in the equation
\begin{equation}
	T=\frac{4\alpha_{a}\alpha_{b}\cos\left(\theta_{a}\right)\cos\left(\theta_{b}\right)}{\alpha_{a}^{2}+\alpha_{b}^{2}+2\alpha_{a}\alpha_{b}\cos\left(\theta_{a}+\theta_{b}\right)}.
	\label{weyl - stept}
\end{equation}

\begin{figure}
	\centerline{\includegraphics[scale=0.4]{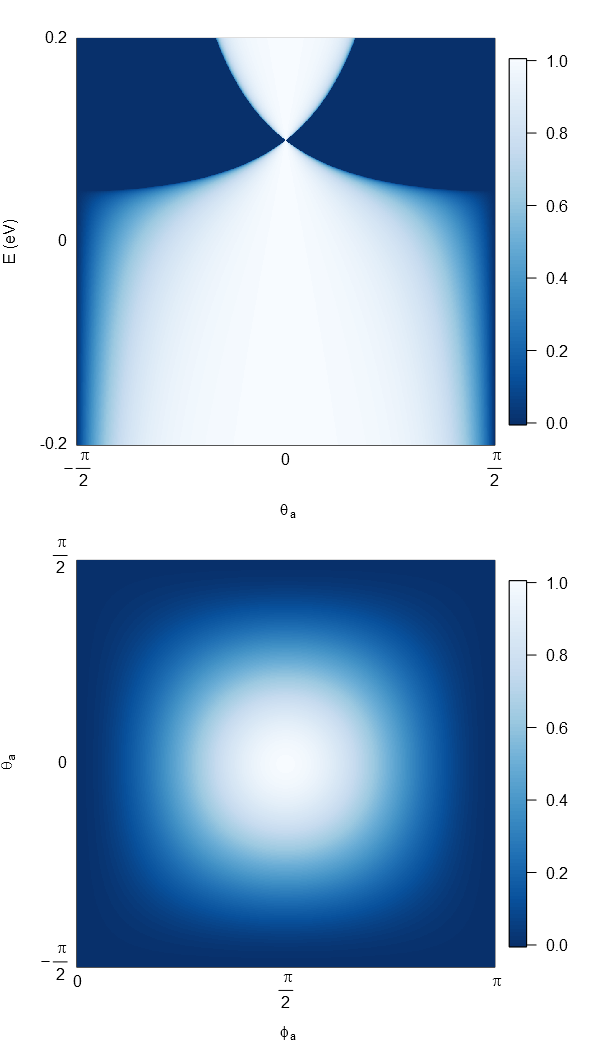}}
	\caption{Density plots for transmission against energy and incident angle for a potential step. The step shown has the heights $V_{a}=0$ eV and $V_{b}=0.1$ eV. (Top) The energy dependence is then shown with $\phi_{a}=\pi/2$. (Bottom) The angular dependence is shown in the step for an energy of $0.05$ eV.}
	\label{weyl-t-step}
\end{figure}

	The full methods used to obtain this have been included in full in the supplementary information\cite{Supplementary}. To obtain the plots in Fig. \ref{weyl-t-step} the correct charge carrier directions must be considered. At $x=0$ there is an electron-hole interface, therefore, the right travelling charge in Fig. \ref{weyl-step} must be carried by a left travelling hole and a right travelling electron. To allow for this change in direction the incident angles of the charge carriers must be changed so that $\theta_{h}=\pi-\theta_{e}$ \cite{b40} where the subscript $e$ and $h$ denote an electron or a hole respectively. With these considerations, the result in Eq.(\ref{weyl - stept}) perfectly recreates the graphene result for a potential step \cite{b13, b52} when the $\phi$ dependence is removed.

	The density plots in Fig. \ref{weyl-t-step} show the symmetry between the $\theta_{a}$ and $\phi_{a}$ angles and transmission that reduces close to the step height. The top plot in Fig. \ref{weyl-t-step} also shows the angular dependent transmission gap. It is very similar to one obtained in graphene \cite{b16} but here it is appearing in the three-dimensional Weyl semimetal heterostructure. Such a heterostructure can be used to focus the electron flow.


\section{Scattering Properties Through a Rectangular Potential Barrier}
\begin{figure}
	\centerline{\includegraphics[scale=0.2]{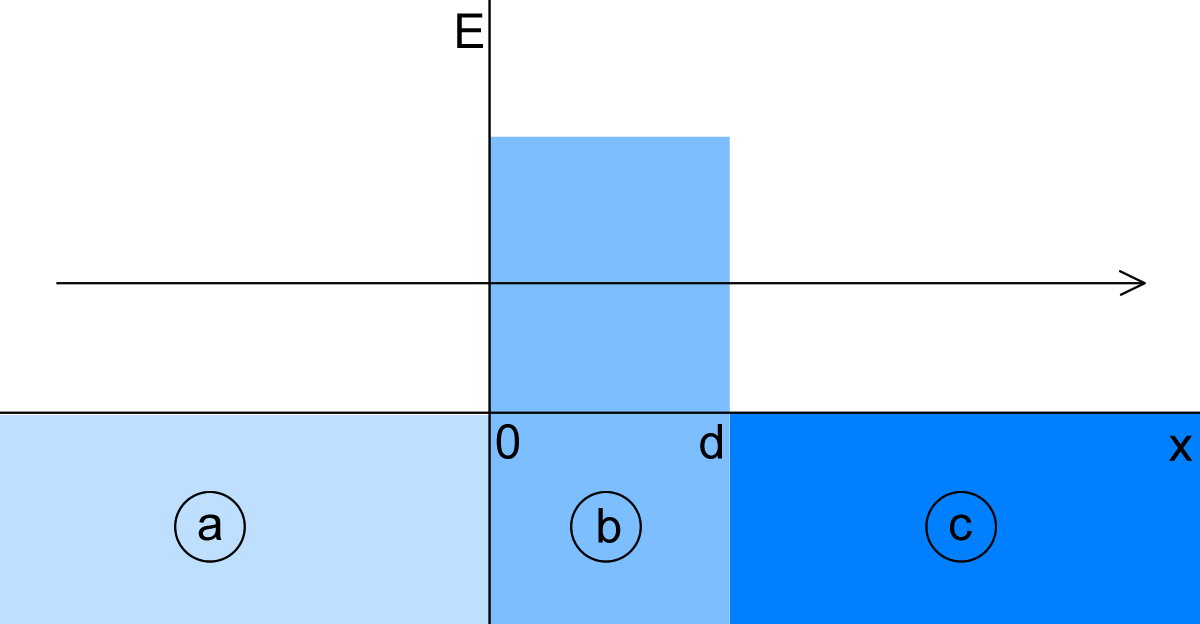}}
	\caption{Diagram of the potential barrier problem. It can be made from two or three different Weyl semimetals by combining two heterostructures similar to the ones demonstrated in Fig. \ref{weyl-step}. A potential barrier is placed in the $x$-direction with a height $V_{b}$ and a width $d$. The shaded region shows where hole transport is present. The three independent regions have been labelled as $a,b$ and $c$ and may correspond, for example, to Weyl semimetals such as NbAs, NbP and TaAs. Such a three layers Weyl semimetal structure can also act as a Veselago lens.}
	\label{weyl-symmetrical-flat}
\end{figure}
	The potential barrier described in Fig. \ref{weyl-symmetrical-flat} is a quasi one-dimensional system with three distinct regions associated with three different materials such as NbAs, NbP and TaAs. An external potential is applied to the center region, creating a scattering system with reflection points at $x=0$ and $x=d$.  If there are different materials used in such a structure the potential steps are naturally created due to the different work functions and the affinity of these materials. At present we are not interested in what creates the potential $V$, just assume that it does exist. If it is possible to control the value of the potential $V$, this will be a Weyl semimetal transistor (for comparison, the design and an operation of similar devices made from graphene\cite{Stone,Hartmann}).

	The wave functions in Eq.(\ref{wave-functions}) can be used in the scattering problem described in Fig. \ref{weyl-symmetrical-flat} and take the form of
	\begin{equation}
		\psi_{a}=
		\left[\begin{array}{ccc}
			\left(a_{1}e^{iq_{a}x}+a_{2}e^{-iq_{a}x}\right)e^{ik_{y}y}e^{ik_{z}z}\\
			\left(a_{1}\alpha_{a}e^{iq_{a}x+i\theta_{a}}-a_{2}\alpha_{a}e^{-iq_{a}x-i\theta_{a}}\right)e^{ik_{y}y}e^{ik_{z}z}
		\end{array}\right],
	\end{equation}
	\begin{equation}
		\psi_{b}=
		\left[\begin{array}{ccc}
			\left(a_{3}e^{iq_{b}x}+a_{4}e^{-iq_{b}x}\right)e^{ik_{y}y}e^{ik_{z}z}\\
			\left(a_{3}\alpha_{b}e^{iq_{b}x+i\theta_{b}}-a_{4}\alpha_{b}e^{-iq_{b}x-i\theta_{b}}\right)e^{ik_{y}y}e^{ik_{z}z}
		\end{array}\right]
	\end{equation}
	and
	\begin{equation}
		\psi_{c}=
		\left[\begin{array}{ccc}
			\left(a_{5}e^{iq_{a}x}+a_{6}e^{-iq_{a}x}\right)e^{ik_{y}y}e^{ik_{z}z}\\
			\left(a_{5}\alpha_{a}e^{iq_{a}x+i\theta_{a}}-a_{6}\alpha_{a}e^{-iq_{a}x-i\theta_{a}}\right)e^{ik_{y}y}e^{ik_{z}z}
		\end{array}\right].
	\end{equation}

	Here regional subscripts have been added to groups of constants and wave function components for left and right travelling waves have been included as required for the transfer matrix method. From the continuity of the wave functions at the barrier interfaces $x=0$ and $x=d$, $\psi_{a}=\psi_{b}$ and $\psi_{b}=\psi_{c}$ respectively. The transfer matrix method \cite{b18} can be used to find the scattering properties through the potential barrier, which is shown in full in the supplementary information\cite{Supplementary}.
	
	Evaluating the transfer matrix allows the transmission coefficient and the total transmission to be obtained. From transfer matrix theory $t=1/M_{2,2}$ and $T=|t|^{2}$ resulting in the equation

\begin{equation}
	T_{weyl}=\frac{4\alpha_{a}^{2}\alpha_{b}^{2}cos^{2}\left(\theta_{a}\right)cos^{2}\left(\theta_{b}\right)}{\beta_{1}+\beta_{2}}
	\label{barrier-t}
\end{equation}
with 
\begin{equation}
 \beta_{1} = 4\alpha_{a}^{2}\alpha_{b}^{2}cos^{2}\left(dq_{b}\right)cos^{2}\left(\theta_{a}\right)cos^{2}\left(\theta_{b}\right)
\end{equation}
and 
\begin{equation}
\beta_{2} = sin^{2}\left(dq_{b}\right)\left(2\alpha_{a}\alpha_{b}sin\left(\theta_{a}\right)sin\left(\theta_{b}\right)-\alpha_{a}^{2}-\alpha_{b}^{2}\right)^{2}.
\end{equation}

\begin{figure}
	\centerline{\includegraphics[scale=0.4]{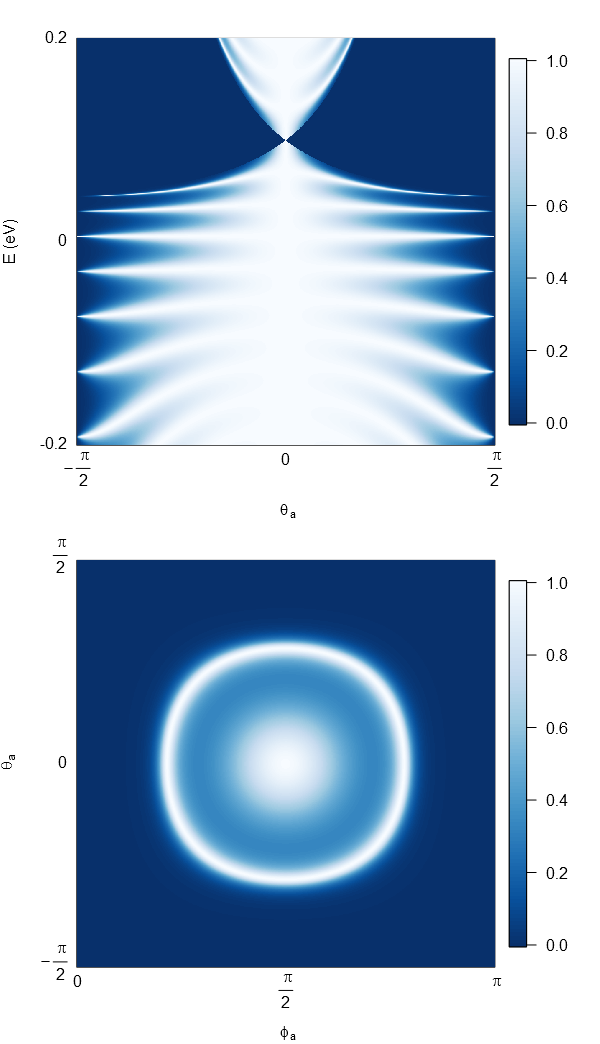}}
	\caption{Density plots for transmission against energy and incident angle with one fixed angle. The potential barrier has a height $V_{b}=0.1$ eV and width $d=100$ nm. (Top) The energy dependence is then shown with $\phi_{a}=\pi/2$. (Bottom) The angular dependence is shown in the step for an energy of $0.05$ eV.}
	\label{weyl-t}
\end{figure}

	The density plots in Fig. \ref{weyl-t} show transmission through a potential barrier. In the bottom plot the transmission is symmetrical for both incident angles and shows high transmission probability when both angles are near incidence. 

	By setting one angle to be normal to the barrier the dependence of the other angle can be examined in further detail. Angle $\phi_{a}$ has been set to $\pi/2$ in the top plot of Fig. \ref{weyl-t} so that the $\theta_{a}$ dependence can be shown. Under this condition the result in Eq.(\ref{barrier-t}) perfectly reduces to the graphene result \cite{b1,b11,b12}. When $\theta_{a}$ is set to zero an identical plot is produced, showing the symmetry between the $\theta_{a}$ and $\phi_{a}$ angles for the potential barrier. Each single-angular dependence recreates the characteristics expected from a two-dimensional graphene-like material.

	These plots again show the angular dependent transmission gap witnessed in graphene \cite{b16} and resonances under the condition of $dq_{b}=n\pi$ \cite{b14} where $n$ is an integer. Importantly, under the condition where $\phi_{a}=\pi/2$ the result in Eq.(\ref{barrier-t}) will perfectly reduce to the graphene result in \cite{b1} including Klein tunnelling \cite{b12} at $\theta_{a}=0$.


\section{Pressure Formation of Non-symmetrical Dirac Cones and Associated Phenomena}
	It was recently established that in the majority of discovered Weyl semimetals, the Dirac cones of opposite chirality are located very close to each other. Their position and orientations must be very sensitive to local lattice distortions which are breaking the spacial inversion. The amplitude of these distortions can be also controlled by applying an external pressure. This pressure may create and change the tilting angle of the Dirac cones or even induce the creation of new pairs of Weyl points. For example, the materials Cd$_{3}$As$_{2}$ \cite{b34,p2,b35} and Na$_{3}$Bi \cite{b29} have recently been shown to possess three-dimensional Dirac cones. The experimental results show a Dirac cone with symmetry in the $k_{x}-k_{y}$ plane, however, in the $k_{x}-k_{z}$ plane there was an asymmetry. This asymmetry may cause these materials to behave differently to symmetrical Dirac cones. In particular the application of the pressure may induce the lattice distortions which break the centre of inversion (CI). Within the kp-method the pressure application can be described by the following Hamiltonian
	\begin{equation}
		 \hat{H}_w=w \vec{d}\cdot \hat{p} \pm v_{f}\hat{p}\cdot\vec{\sigma} 
	\end{equation}
where the vector $\vec{d}$ is related to a vector of atomic displacements induced by the external hydrostatic or uniaxial pressure $P$ and the parameter $w$ describes the tilting of the Weyl cones. According to Hooke's law, these displacements may be described by the elasticity equation, which provides a relationship between stress $\sigma_{s}$ and strain $\varepsilon$:  $\sigma_{s}= K\varepsilon$, where $K$ is known as the elastic modulus or Young's modulus. Taking into acount that the stress $\sigma_{s} \sim P$ and the strain $\varepsilon\sim |\vec{d}|/w$, we have a relation between the external pressure and the vector $\vec{d}$ of the form $ K \nabla\cdot \vec{d} \approx P$. 

\begin{figure}
	\centerline{\includegraphics[scale=0.33]{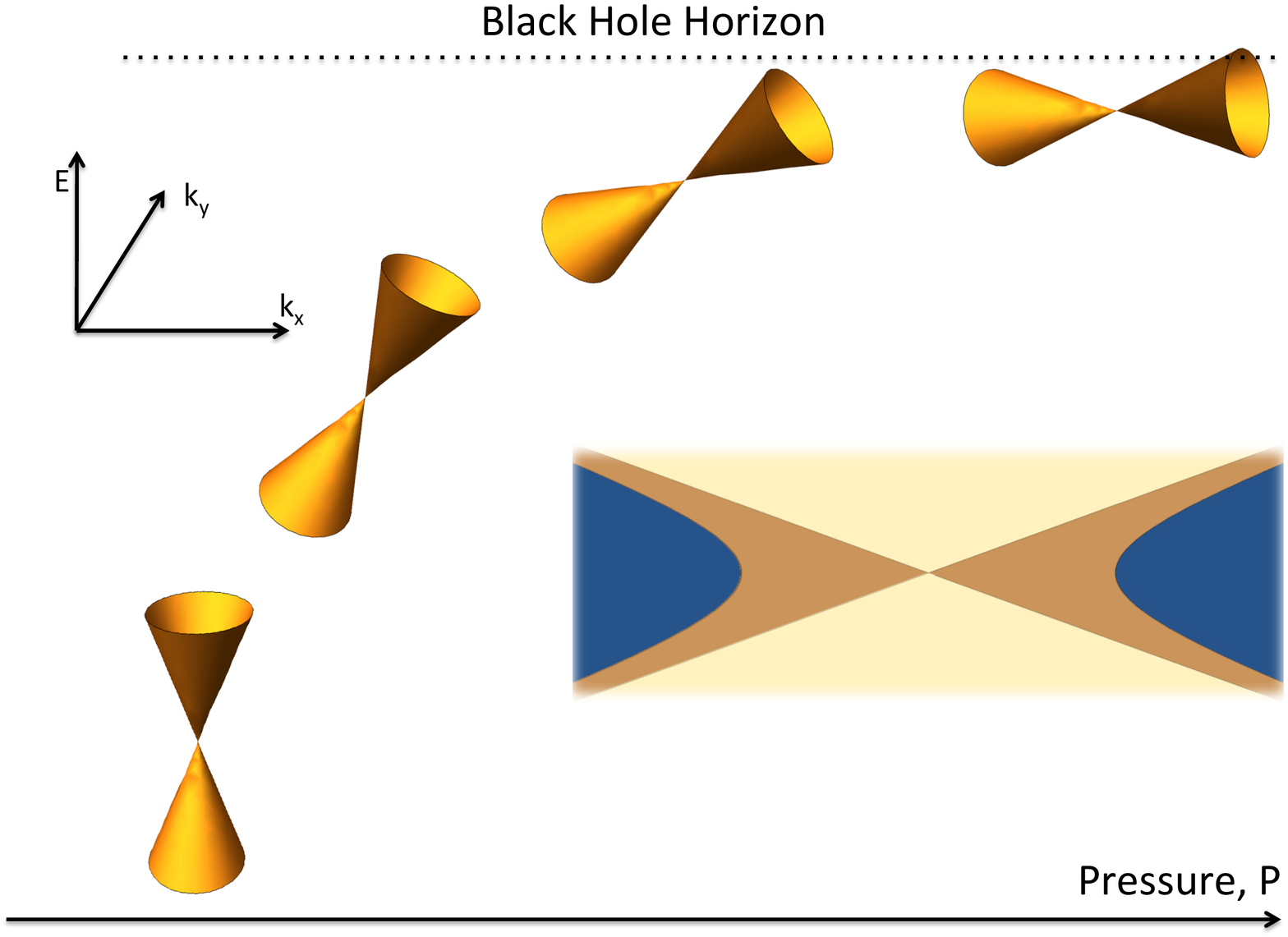}}
	
	\caption{ A schematic diagram showing a deformation of the electronic spectrum of Weyl semimetal under pressure P; the tilting angle of the Dirac cone increases and it finally takes a horizontal position, i.e. tips over. Such tilting of the cone corresponds also to the dramatic changes in the shape and topology of Fermi surface. If at a small angle the Fermi surface has a closed shape of a deformed circle or sphere then at the tipped stage, the Fermi surface is open and consists of two separate parts, see the dark blue areas on the insert in the centre of the Figure. The character of the electronic transport undergoes a dramatic transformation from ballistic and isotropic to diffusive and very anisotropic. Such a transformation of the electronic spectrum in tilted Weyl semimetals strongly resembles the behavior of a light cone when a probe particle is approaching a black hole horizon, where space and time coordinates are interchanged \cite{Novikov-1963}.}
	\label{pressure-weyl}
\end{figure}

\begin{figure}
	\centerline{\includegraphics[scale=0.4]{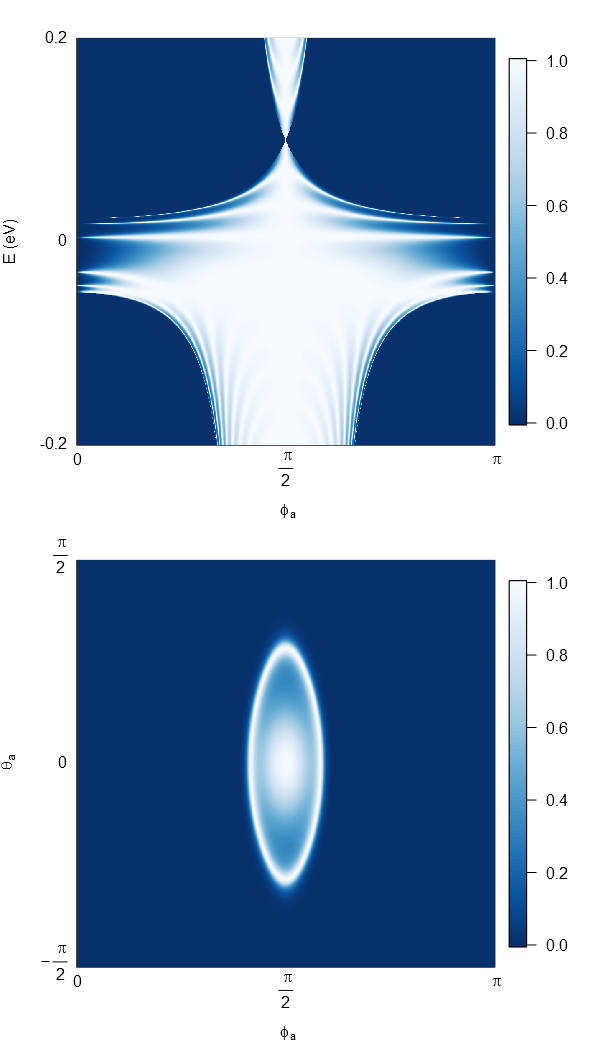}}
	\caption{Density plots for transmission probability against energy and incident angle from Eq.(\ref{barrier-t}) with the modified $k_{z}$ from Eq.(\ref{kz-l}). The potential barrier shown has the characteristics $V_{a}=0$ eV, $V_{b}=0.1$ eV and $\lambda = 1/3$. (Top) The energy dependence is then shown with $\phi_{a}=\pi/2$. (Bottom) The angular dependence is shown in the step for an energy of $0.05$ eV.}
	\label{barrier-l3}
\end{figure}

	In addition, to model the asymmetry of the Dirac cones around the main symmetry axis a scaling factor $\lambda$ can be introduced to $k_{z}$ so that $k_{z} \rightarrow \lambda k_{z}$. Therefore to describe tilted asymmetric Weyl cones we use the following energy momentum relation
		\begin{equation}
			E(k_x,k_y,k_z)=w \hbar \vec{d}\cdot \vec{ k}\pm v_{f}\sqrt{k_{x}^{2}+k_{y}^{2}+\lambda^{2} k_{z}^{2}}
			\label{energy-momentum-l}
		\end{equation}
		and the wave vector, $k_{z}$, may be here defined with the use of the equation:
		\begin{equation}
			k_{z}=\frac{1}{\lambda}\frac{|E-w \hbar \vec{d}\cdot \vec{ k} |}{\hbar v_{f}}\cos\phi.
			\label{kz-l}
		\end{equation}
	The tilting of Weyl cones is described by the parameters $w$ and $d_i$, which can be controlled by the pressure, $P$, (see Fig. \ref{pressure-weyl}). When the pressure $P$ increases, the amplitude of the vector $ \vec{d}$ increases which also increases the tilting angle until the Dirac cone tips. This tilting and the deformation  of Dirac cones described by parameter $\lambda$ (see the change in definition of $k_{z}$ in Eq.(\ref{kz-l})) will affect the scattering and transport properties of any device constructed from these novel materials. For an illustration we show the plots in Fig. \ref{barrier-l3} of the transmission probability for a potential barrier from Eq.(\ref{barrier-t}) with $\lambda=1/3$. The scaling factor $\lambda$ causes the regions of high transmission to reduce and new regions of no transmission are introduced, the tilting of the cones further decreases the transmission probability. The resonances that usually occur in the potential barrier become condensed into the reduced regions of transmission. In Fig. \ref{barrier-l3} the symmetry of resonances is broken, no longer showing circular resonances but a combination of resonance lines and ovals. The drastic decreasing in the transmission is related to a breaking of particle-hole symmetry which arises with the tilting of the cones.

	It is interesting to note that the tilting of these Dirac cones may model matter at extreme conditions. The symmetric Dirac cone represents a conventional relativistic light cone which describes all events in the past, at present and in the future, following a conventional causal relation. However, when approaching a black hole and in a vicinity of the black hole horizon, time is slowed. Eventually, at the black hole horizon time is stopped, that is it is swapped with the spacial coordinate. At this moment the light cone is tipped. This was first noticed by Igor Novikov\cite{Novikov-1963,Novikov-1989,Novikov-1989a}, who used the effect of the space-time interchanging to describe the possibility of travelling in time and to built up a first hypothetical time machines. More precisely, for time travelling we need a worm-hole but it is irrelevant for the present discussion here (see, for details\cite{Novikov-1990}).

	The inter-changing of space and time corresponds to a horizontally tilted position of the light cones; here in Weyl semimetals the horizontal tilting or the effect of the 'black hole' can be achieved by the application of an external hydrostatic or uniaxial pressure. Under these conditions the Fermi surface opens and alters the quasiparticle trajectories; the character of the transport changes from ballistic (which is still valid at small tilting angles) to highly diffusive. Thus by applying a pressure to the Weyl semimental having no CI one may model a matter at extreme conditions arising in vicinity of the black hole horizon.

\section{Photo-current and Light Absorption in Hyperbolic Dirac Phase}
	Now we are in the position to characterise the state associated with tipped Dirac cones; the hyperbolic shape of the Fermi surface is similar to that of a hyperboloid mirror which provides a unique possibility for electron transport. When an electrical field is applied and the electron momenta changed, electrons may collect in the focal points of the Fermi surface such as in hyperbolic mirrors. This effect needs to be further elaborated.

	Another point is a strong and anisotropic light absorption. As an illustration and for a simplicity we consider an interaction of electrons in a tipped two-dimensional Dirac cone, $E(k_x,k_y)=v_f \hbar \sqrt{k_x^2-k_y^2}$  with light that is linear polarised i.e. $ \vec{E}=(E_x \cos {\omega t}, E_y \cos {\omega t})$. In a simple quasi-classical approximation with a single relaxation time approximation, $\tau$, light induces the photo-current, with a value that can be estimated with the use of the time-dependent path integral approach as a steady solution of the time-dependent Boltzmann transport equation (for details see Refs \cite{Budd,Mac,Hramov-2014}). This calculation gives the equation for the dependence of the current on the amplitude of the electromagnetic field, $(E_x , E_y)$
\begin{equation}
 \vec{j}(E_x, E_y)=\frac{e n_0 v_f}{\sqrt{E_x^2-E_y^2}}  (E_x , E_y)
\end{equation}
where $n_0$ is the electron density, which is assumed small. This expression is valid when $ E_x\neq E_y$. The singularity vanishes when we take into account  fluctuations and next orders of the perturbation theory. Nonetheless the obtained expression indicates the strength and the anisotropy of the photoeffect, which can be observed here. The light absorption, which is defined as an average $w=< \vec{j}\cdot\vec{E}>$ gives the following expression for the light absorption
\begin{equation}
w=\frac{e n_0  v_f}{1+\omega^2\tau^2}  \frac{ |\vec{E}|^2}{\sqrt{E_x^2-E_y^2}}. 
\end{equation}

	Again we see, as in the case of the photo-current, the effect is huge when the polarisation of the light is oriented along the rays of the Dirac cone, limited by the conditions $k_x=k_y$ or $E_x=E_y$. This singularity in the photocurrent and light absorption is exactly related to the hyperbolic character of the electronic spectrum. Thus, the new hyperbolic phase when the Dirac cone is tipped has unique properties; highly anisotropic enhanced photo-effect and light absorption.



\section{Comparison with Optics }

	The linear spectrum of Weyl fermions provides an opportunity for comparison with conventional optics. Unlike an electromagnetic wave the transmission properties of Weyl fermions cannot be entirely separated in the spacial dimensions. In the transmission probability through a barrier for Weyl fermions (Eq.(\ref{barrier-t})) there is a phase factor from the second wave function component not found in the transmission probability for an electromagnetic wave
	\begin{equation}
		T_{optics}=\frac{4k^{2}q^{2}}{4k^{2}q^{2}\cos^{2}\left(qd\right)+\left(k^{2}+q^{2}\right)^{2}\sin^{2}\left(qd\right)}.
	\end{equation}

	The derivation and definitions for this can be found in the supplementary information\cite{Supplementary}. The similarities between $T_{weyl}$ and $T_{optics}$ are most obvious when considering that both systems experience resonance under the condition $dq=n\pi$. The resonance condition for the three-dimensional Weyl fermions expands to
	\begin{equation}
		E_{weyl}=V_{b}\pm\hbar v_{f}\sqrt{\frac{n^{2}\pi^{2}}{d^{2}}+k_{y}^{2}+k_{z}^{2}}
	\end{equation}
	and in the optical case
	\begin{equation}
		E_{optics}=\frac{\hbar cn\pi}{dn_{b}}
	\end{equation}
	where $n_{b}$ is the refractive index of the intermediate medium. From the optical rule $q=nk$ the 'refractive index' of Weyl fermions travelling though a potential will be of the form
	\begin{equation}
		n_{weyl}=\sqrt{\frac{\left(E-V\right)^{2}-\hbar^{2} v_{f}^{2}\left(k_{z}^{2}+k_{y}^{2}\right)}{E^{2}-\hbar^{2} v_{f}^{2}\left(k_{z}^{2}+k_{y}^{2}\right)}}.
	\end{equation}

	However, from the probability current it is known that the transport through a potential barrier for Weyl fermions is band to band. For this reason at energies below the barrier $(E<V)$ a left travelling charge will require a left travelling electron and a right travelling hole. To represent this the refractive wave number comparison must use a different component of the wave function so that $q=-nk$, resulting in a new refractive index inside the barrier
	\begin{equation}
		n_{weyl}=-\sqrt{\frac{\left(E-V\right)^{2}-\hbar^{2} v_{f}^{2}\left(k_{z}^{2}+k_{y}^{2}\right)}{E^{2}-\hbar^{2} v_{f}^{2}\left(k_{z}^{2}+k_{y}^{2}\right)}}
	\end{equation}
	which will be negative for all values. Using the rules for electromagnetic waves (stated in the supplementary information\cite{Supplementary}) a wavelength for the Weyl fermions may be expressed as
	\begin{equation}
		\lambda_{weyl}=\frac{2\pi}{k}=\frac{2\pi}{\sqrt{\frac{\left(E-V\right)^{2}}{\hbar^{2}v_{f}^{2}}-k_{y}^{2}-k_{z}^{2}}}
	\end{equation}

	With the expressions for positive or negative refractive indices and wavelength it is possible to apply further optical analysis to the charge carriers in a Weyl semimetal. It was theoretically shown that graphene p-n junctions may focus electron flow as the two-dimensional analogue of the Veselago lens\cite{p1} as expected from situations where a negative refractive index arises.


\section{Scanning Tunnelling Microscope with a Veselago Lens as a Probing Tip}		

\begin{figure}
	\centerline{\includegraphics[scale=0.33]{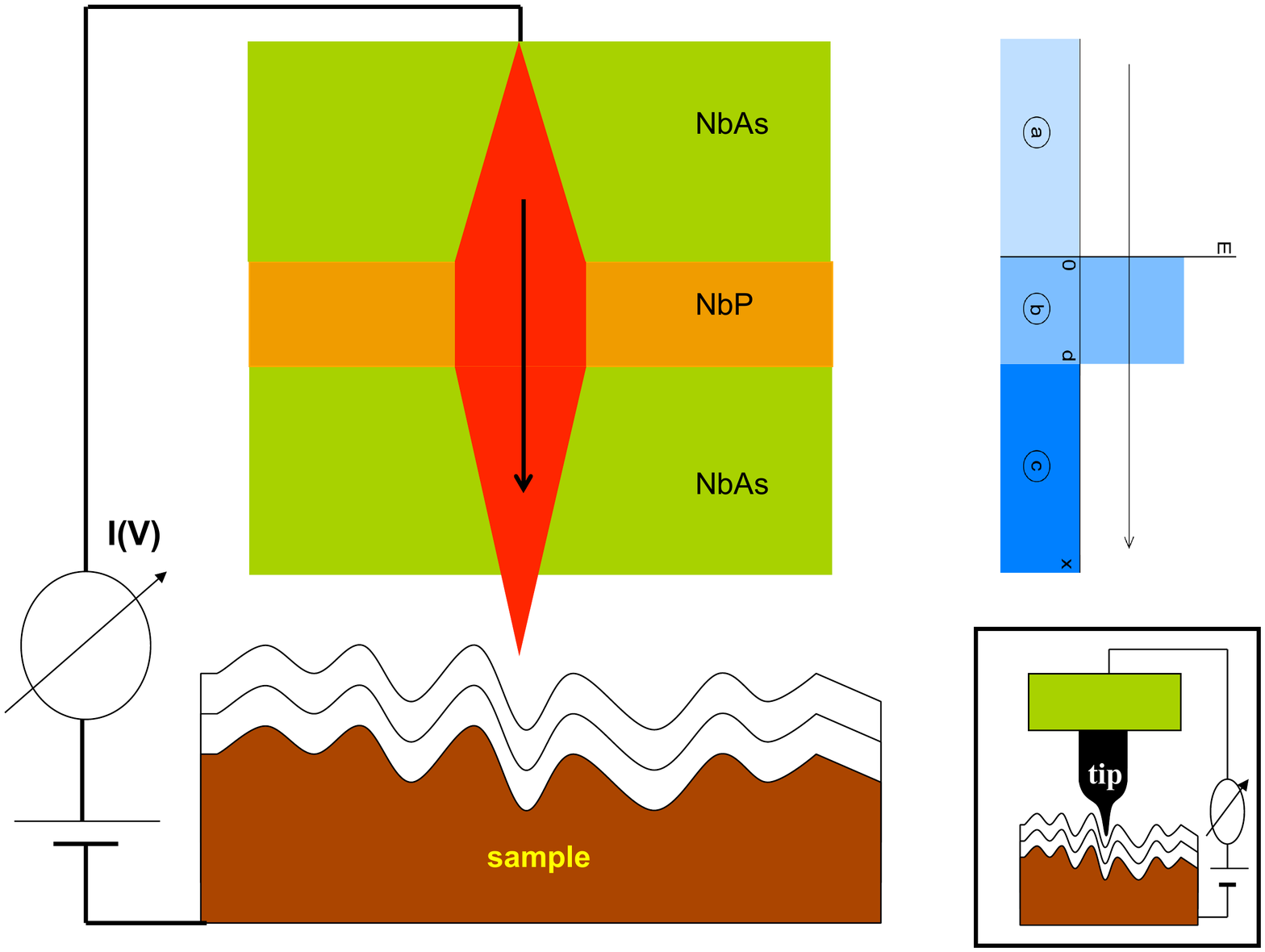}}
	\caption{A schematic operation of a scanning tunnelling microscope (STM) with probing tip built from Weyl semimetals (given in green and orange). It is made in the form of a Veselago lens, e.g. in the form of three layers of NbAs and NbP as given in the main pane. In this case, the electron propagation through these three layers act as a rectangular potential barrier (shown in blue on the right side). The electron flow supplied by the electrical current to the top layer is focused on a small spot in the bottom layer (given in red). Due to the properties of the Veselago lens associated with the materials negative refractive index, the focusing spot for electrons can be very small and may increase the spatial and temporal resolution of the STM. By measuring the current-voltage characteristics $I(V)$ and scanning the surface one may in principle, identify various electron clouds, atomic orbitals and resolve the existing issues with chemical bonding and kinetics.}
	\label{tip}
\end{figure}

	The negative refractive index for electron waves in Weyl semimetals allows for the creation of a three-dimensional Veselago lens, which can be used as a probing tip in a scanning tunnelling microscope (STM). A diagram of such an STM is shown in Fig. \ref{tip}. A structure made from three layers of different Weyl materials (given in green and orange in Fig. \ref{tip}) may focus a three-dimensional flow of electrons as light is focused in an optical Veselago lens. In this three layer structure the energy position of the Weyl points associated with the middle layer (shown in orange) will be above or below the Weyl points in the top and bottom layers (shown in green). The structure acts as a rectangular potential barrier for electron propagation (shown in blue on the right side) as discussed in previous sections. In this case if  we apply an electron current to the top layer the system may focus electrons on the bottom layer. The focusing flow is shown in red. Such a lens may be very useful in STMs, which allows for the study of the atomic structure of materials using the materials current voltage characteristics $I(V)$, measured with the electrical circuit given in Fig.\ref{tip}; an STM with a Weyl semimetal tip is schematically presented on the main pane of the Figure. For comparison, the insert in the right lower corner of the Figure is a conventional STM with sharp atomic tip (in black). Currently, the STM is one of the most powerful imaging tools available; they can image materials and atoms on their surface with an unprecedented precision.  At present a good resolution for an STM is considered to be 0.1 nm lateral resolution and 0.01 nm depth resolution\cite{Bai}. However to get a good resolution for these images we have to use atomically sharp probing tips; creating these remains something of a dark art and not easily achieved and is, definitely, not in a mass production.
 		 
	Note that the propagation of electrons should be ballistic in the semimetals used for the Weyl probing tip. This condition arises when the energy of electrons is in the vicinity of the Weyl points and these electrons are acting as light. Due to the gapless linear Dirac spectrum, the wavelength of the electrons in the vicinity of these Weyl points is divergent meaning they are acting as waves with a very large wavelength. Therefore they will have little scattering from impurities or defects and propagate ballistically (see the discussion at the end of the original paper\cite{b2}). If we make tip for an STM in the form of three layers of Weyl semimetals (as shown in Fig. \ref{tip}) we can significantly increase the performance of the standard STM and in particular its resolution. The latter arises because of the perfection and small size of the focusing spot formed by the Veselago lens\cite{p1}.

	With the new STM probing tip made from Weyl semimetals one may see not only individual atoms within the studied materials but also electron clouds within the atoms. With such Weyl STM we expect that chemical bonds will be routinely imaged and manipulated. Another issue of STM is a temporal resolution, which is typically of the order of seconds. This prevents STMs from imaging the fast kinetics of electrochemical processes. With Weyl semimetal tips one may increase the amplitude of the tunnelling current while keeping the electrons focused on the same small spot on the probing tip surface. With an increased tunnelling current amplitude we expect that this issue with temporal resolution will be also resolved.

\section{Density of States}
			The density of states can be calculated by using the general formula for density of states \cite{bb3}
			\begin{equation}
				\rho\left(E\right)=\sum_{k}\delta\left(E-E_{k}\right)
			\end{equation}
			where $\delta\left(x\right)$ is the Dirac delta-function. Converting sum notation to integration over all momentum
			\begin{equation}
				\rho\left(E\right)=\frac{L_{x}L_{y}L_{z}}{8\pi^{3}}2\int_{k}\int_{\theta}\int_{\phi}\delta\left(E-E_{k}\right)k^{2}dkd\theta d\phi
			\end{equation}
			with $L_{x,y,z}$ being the size of the system in the respective dimension. Using the linear spectrum of Weyl fermions this can be converted to integration over energy
			\begin{equation}
				E_{k}=\hbar v_{f}k,
				\hspace{0.5cm}
				dE_{k}=\hbar v_{f}dk,
				\hspace{0.5cm}
				k^{2}dk=\frac{E_{k}^{2}}{\hbar^{3}v_{f}^{3}}dE_{k}.
			\end{equation}
			Finally by using the integration rule $\int f(x)\delta(x) dx=f(0)$ the density of states becomes
			\begin{equation}
				\rho\left(E\right)=\frac{L_{x}L_{y}L_{z}}{\pi\hbar^{3}v_{f}^{3}}E^{2}
				\label{weyl-dos}
			\end{equation}
			Unlike in graphene, which has a linear density of states, the density of states for a three-dimensional Weyl semimetal is parabolic due to the additional $k_{z}$ component.


\section{Landauer Formalism}

\begin{figure}
	\centerline{\includegraphics[scale=0.25]{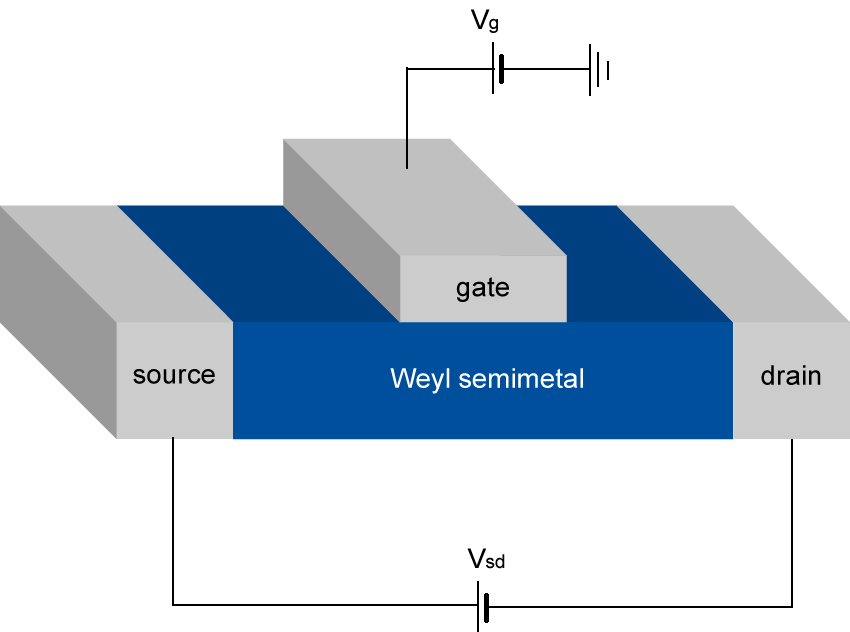}}
	\caption{A simple example of a Weyl semimetal transistor with current-voltage characteristics described by Eq.(\ref{weyl-i}).}
	\label{transistor}
\end{figure}

\begin{figure}
	\centerline{\includegraphics[scale=0.4]{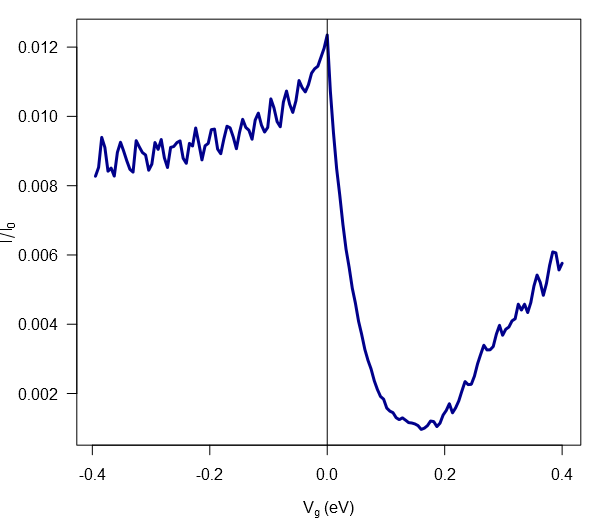}}
	\caption{The gate voltage ($V_{g}$) dependence on current for a Weyl semimetal transistor from Eq.(\ref{weyl-i}) with $I_{0}=e\frac{2L_{y}L_{z}}{\pi\hbar^{3}v_{f}^{2}}$, $V_{b}=V_{g}$, $V_{sd}=0.2$ eV, $d=100$ nm and $T=298$ K.}
	\label{barrier-ivg}
\end{figure}

\begin{figure}
	\centerline{\includegraphics[scale=0.4]{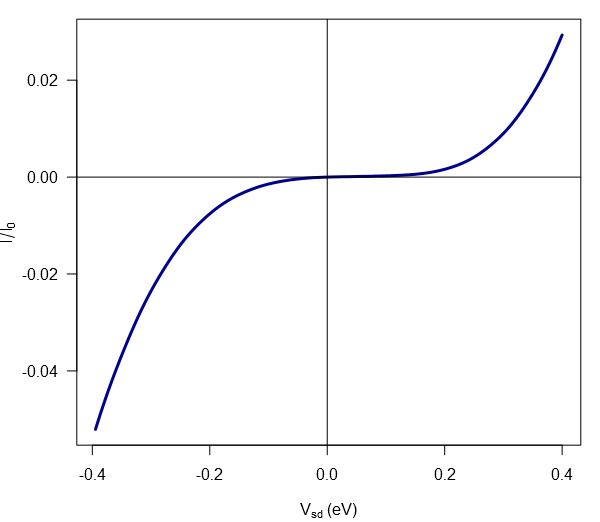}}
	\caption{The source-drain voltage ($V_{sd}$) dependence on current for a Weyl semimetal transistor from Eq.(\ref{weyl-i}) with $I_{0}=e\frac{2L_{y}L_{z}}{\pi\hbar^{3}v_{f}^{2}}$, $V_{b}=0.1$ eV, $d=100$ nm and $T=298$ K.}
	\label{barrier-isd}
\end{figure}

\begin{figure}
	\centerline{\includegraphics[scale=0.4]{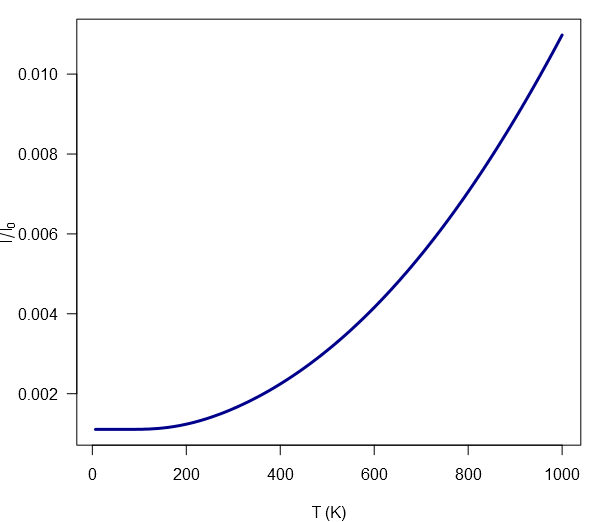}}
	\caption{The temperature dependence on current for a Weyl semimetal transistor from Eq.(\ref{weyl-i}) with $I_{0}=e\frac{2L_{y}L_{z}}{\pi\hbar^{3}v_{f}^{2}}$, $V_{b}=0.1$ eV, $d=100$ nm and $V_{sd}=0.2$ eV.}
	\label{barrier-it}
\end{figure}

	The current through the scattering systems formulated earlier can be calculated with the Landauer formalism for ballistic transport. In this model perfect electron emitters are connected to a scattering device via perfectly conducting wires. The electron emitters emit electrons up to the quasi-Fermi-energy $\mu_{L}$ and $\mu_{R}$ into the respective side of the scattering device. In this model the current through the scattering device is given in Ref.\cite{b6} as
\begin{equation}
	I=ev_{f}\frac{dn}{dE}T\left(\mu_{L}-\mu_{R}\right)
	\label{current-dos}
\end{equation}
where $e$ is the electron charge, $v_{f}$ is the Fermi velocity and $dn/dE$ is the density of states. At a finite temperature the electron emitters inject electrons as described by the Fermi-Dirac distribution
\begin{equation}
	f_{L,R}=f\left(E-\mu_{L,R}\right)=\frac{1}{e^{\frac{E-\mu_{L,R}}{k_{b}t}}+1}
\end{equation}
instead of up to the quasi-Fermi-energies $\mu_{L}$ and $\mu_{R}$. Here $k_{b}$ is the Boltzman constant and $t$ is the temperature. Using the density of states for Weyl fermions in Eq.(\ref{weyl-dos}) and integrating over energy and incident angle produces the $x$-direction current

		\begin{equation}
			\frac{I_{x}}{I_{0}}=\int^{\infty}_{-\infty}\int^{\frac{\pi}{2}}_{-\frac{\pi}{2}}\int^{\pi}_{0}T\left[f_{L}-f_{R}\right]E^{2}\cos\left(\theta\right)\sin\left(\phi\right)dEd\theta d\phi
			\label{weyl-i}
		\end{equation}
with the group of constants $I_{0}=e\frac{2L_{y}L_{z}}{\pi\hbar^{3}v_{f}^{2}}$ and $L_{y,z}$ is the length of the system in the respective direction. Additional derivations can be found in the supplementary information\cite{Supplementary}. A suitable device has been suggested in Fig. \ref{transistor}; the characteristics of such a device can be seen in Fig. \ref{barrier-ivg}-\ref{barrier-it}.

	The plot of current against barrier height in Fig. \ref{barrier-ivg} shows oscillations and a sharp drop in current between $0$ and $0.2$ eV. The drop in current is caused by the low point in transmission (shown at $E=V_{g}$ in Fig. \ref{weyl-t}) coinciding with the energy region contained with the Fermi-Dirac distributions $f_{L}-f_{R}$. This will happen when $V_{g}\approx eV_{sd}$, outside of this region the Fermi-Dirac distributions will be centered around high transmission regions and the current will increase, as shown when $|V_{g}|>V_{sd}$.

	Fig. \ref{barrier-ivg} shows a similar current-voltage curve to a Zener diode with a slight bias to positive voltage. This is again due to the low point of transmission coinciding with the region contained by the Fermi-Dirac distributions. The bias can be changed to negative voltage by setting $V_{g}\rightarrow -V_{g}$, however the plateau caused by the low transmission may be moved by increasing the magnitude of the barrier.

	The current steadily increases with temperature in Fig. \ref{barrier-it}, showing a parabolic dependence. As the temperature increases, less of the ballistic charge carriers will be within the energy range of the potential barrier, as fewer of the charge carriers are scattered by the barrier; the current increases.

\section{Conclusion}

	Here we have identified the scattering properties of a three-dimensional material Weyl semimetals with a linear dispersion relation such as; NbAs, NbP, TaAs, TaP, Ag$_{2}$Se or Cd$_{3}$As$_{2}$. Specifically we focused on the exact solution of tunneling of Weyl or Dirac fermions through a potential barrier and potential steps. This is different from two-dimensional Dirac electrons existing in graphene and itself a very important and fundamental problem in the view of a large number of different Weyl semimetals discovered in recent years. To describe the properties of these materials we have used a three-dimensional two by two Weyl Hamiltonian describing one of two Weyl points (WPs), which always exist in pairs. Each of these points represents a topological defect of the band structure  and has a shape of the Dirac monopole. In fact, these two Weyl points correspond to the monopole (a positive topological charge) and to the anti-monopole (a negative topological charge), respectively which will always appear and disappear in pairs. The fermions associated with each Weyl point have a fixed chirality corresponding to the topological charge. When there is no magnetic field included in the electron transport the chirality is preserved and therefore it is possible to study the transport of the Weyl fermions using one effective two by two Weyl Hamitonian. 
	
	The situation will be different when a magnetic field is applied parallel to the applied electrical field. A parallel magnetic field creates a chiral anomaly, where the number or total electrical charge of the fermions with one chirality increases, while the fermions with the other chirality decreases. Such an anomaly is of course related to the fundamental symmetries; the time reversal, T and the spacial inversion, P. Without the magnetic field, these symmetries hold, for example, P,  applied individually may transform one Weyl point into the other or one chirality into the other. One may find a unitary transformation which transforms one point into the other and therefore the electron transport in a weak electrical field associated with these points is chirality independent and, therefore, additive. When the electrical or magnetic field is applied separately the symmetries,  P or T break respectively.  However when the magnetic field is parallel to the electrical field, the symmetry transformation associated with the product, PT, remains invariant and, therefore, the chiral anomaly arises\cite{anomaly-1983}. There the particle number for a given chirality is not conserved quantum mechanically. This phenomenon is also known as the Adler-Bell-Jackiw anomaly. We have not studied this case here, which also should involve surface states of the Weyl semimetals; further studies are needed, which we will do in the future.

	In the most of the existing cases WPs appear because of the lack of an inversion center in the crystal structure. In TaAs there are twelve pairs of Weyl points, they arise in the vicinity of bands crossing which occur along main mirror-invariant planes. When the spin-orbital interaction (SO) is taken into account the band crossing is transformed into WPs. These WPs appear as isolated gapless nodes slightly off the symmetry plane. Two of the pairs of Weyl points in TaAs are located close to the Fermi energy (about 2 meV above it and near the $k_z=0$ plane) in the centre of the Brillouin zone. The remaining Weyl points are about 21 meV below the Fermi level. They form sixteen electron pockets providing a very small electron density in this many valley system. In this case the scattering properties are defined by the two pairs of  WPs which are about 2 meV near the Fermi surface. Due to the chirality conservation for the scattering properties we have considered each of these points separately.

	Moreover, we have also discussed the properties of type two Weyl semimetals, which have a tilted Dirac cone-like structure around the band touching points. We show that this tilting may be controlled by an external pressure and that the changing of the titling angle will lead to various novel physical properties of Weyl semimetals. In particular the transmission properties through the potential step or barrier depend strongly on the shape and titling of the Dirac cones. When a strong  external hydrostatic or uniaxial pressure is applied the Dirac cones can be tipped over. The tipping over of the Dirac cones will cause a new hyperbolic Dirac phase, where  the Fermi surface has a hyperbolic open shape consisting of two separate parts resulting in electron transport that is very anisotropic and very diffusive. We show that in this case there may arise new phenomena such as a huge anisotropic photocurrent generation and large anisotropic light absorption. The open hyperbolic shape of the  Fermi surface may lead to some new resonance phenomena in a magnetic field, which may be associated with the focal points (in momentum space)  of this hyperboloid structure. We also draw an analogy between the tilting of the Weyl cones (arising under pressure) and an analogous phenomenon arising when a particle approaches a black hole. In the latter case a tilting of the light cone of spacetime occurs, where the analogous 'Weyl point' is separating events in the future from events in the past. The infinitely strong gravitational field on the black hole horizon is forces the light cone to tip over\cite{Novikov-1963}; the time and spacial coordinates are interchanged and conditions for time traveling arise\cite{Novikov-1990}. It is interesting that we may have analogous phenomena in Weyl semimetals which may be stimulated by external pressure.

	We have shown a strong analogy between Weyl fermions and photons propagating in media with a negative refractive index. A negative refractive index will then allow a Weyl semimetal to behave as a metamaterial. Thus three-dimensional bulk Weyl semimetals can be used to make a Veselago lens. As an example, a lens can be created from three layers of Weyl semimetal films deposited on each other in a sequence, such as NbAs-Nb-P-NbAs. This structure forms a rectangular potential barrier for the Weyl fermions propagating in the direction perpendicular to the films. The Veselago lens can focus electrons on a very small spot with high efficiency and can be used for the probing tip in a scanning tunneling microscope (STM).  In Fig.\ref{tip} we presented a simplified schematic for the operation of an STM created with a Weyl semimetal tip in the form of a Veselago lens. It is presented by three layers of Weyl semimetals given in green and orange. This resolution depends on the size of the focused electron spot and can be further improved by proper focusing of the electron motion. Therefore such Weyl-Veselago lens can significantly improve the resolution of the present STM technology and we expect that this proposal will be realised very shortly.

	It is also very important to note that now 'Weyl semimetals' and other topological materials have also been discovered in nano-photonics and metamaterials\cite{Ling-2013,Ling-2015}. The Weyl points have been found in the structures which were artificially designed. The experimental discovery of the photonic Weyl points pave the way to topological photonics, where real Veselago lenses will be made. We now have a very exciting time for the realisation of optical metamaterials and a new imaging technology knocking in the door. This includes a range of new opportunities throughout photonics, e. g. where interfaces may support new states of light as well as unidirectional waveguides that allow light to flow around large imperfections without back-reflection\cite{Tena-2015}.

	The results obtained can be equally used in Weyl semimetals and topological nanophotonics. In particular, the transmission properties through a one-dimensional potential step show angular symmetry between $\theta_{a}$ and $\phi_{a}$ at all energies. This circular symmetry of the electron beam in Weyl semimetals gives rise to the electron angular momentum which is similar to the polarisation of light. So we may have 'linear' or 'circular polarisation' of electrons; a similar situation arises for the Weyl fermions penetration through a potential barrier. Of course our results can be mapped to a lower dimensional case as in graphene.  For example, the reduced result with $\phi_{a}=\pi/2$ perfectly recreates the two-dimensional graphene result and features similar properties such as angle dependent transmission gap, Fabry-P\'{e}rot resonances and Klein tunnelling. With the transmission probability and the Landauer formalism, the ballistic current through a Weyl transistor was predicted with the dependencies on barrier height, source-drain voltage and temperature. The current voltage characteristics show similar properties to both conventional electronics as well as to other linear spectrum materials such as graphene. The work here aims to further the understanding of gapless semiconductors and highlight three-dimensional materials with graphene-like properties, resulting in a guideline for the properties expected from Weyl semimetal devices.

\end{document}